# Constraining the shape of a gravity anomalous body using reversible jump Markov chain Monte Carlo

**Xiaolin Luo**

CSIRO Mathematical and Information Sciences, Sydney, Locked bag 17, North Ryde, NSW, 1670, Australia.

E-mail: Xiaolin.Luo@csiro.au

**Abstract.** Typical geophysical inversion problems are ill-posed, non-linear and non-unique. Sometimes the problem is trans-dimensional, where the number of unknown parameters is one of the unknowns, which makes the inverse problem even more challenging. Detecting the shape of a geophysical object underneath the earth surface from gravity anomaly is one of such complex problems, where the number of geometrical parameters is one of the unknowns. To deal with the difficulties of non-uniqueness, ill-conditioning and nonlinearity, a statistical Bayesian model inference approach is adopted. A reversible jump Markov chain Monte Carlo (RJMCMC) algorithm is proposed to overcome the difficulty of trans-dimensionality. Carefully designed within-model and between-model Markov chain moves are implemented to reduce the rate of generating inadmissible geometries, thus achieving good overall efficiency in the Monte Carlo sampler. Numerical experiments on a 2-D problem show that the proposed algorithm is capable of obtaining satisfactory solutions with quantifiable uncertainty to a challenging trans-dimensional geophysical inverse problem. Solutions from RJMCMC appear to be parsimonious for the given prior, in the sense that among the models satisfactorily represent the true model, models with higher posterior probabilities tend to have fewer number of parameters. The proposed numerical algorithm can be readily adapted to other similar trans-dimensional geophysical inverse applications.

**Revised version: November 27, 2009**

# 1. Introduction

Monte Carlo techniques for geophysical inversion were first used about forty years ago (Keilis-Borok and Yanovskaya 1967; Press 1968, 1970, Anderssen and Seneta 1971, Anderssen et al 1972), since then there has been considerable advances in both computer technology and mathematical methodology, and therefore an increasing interest in those methods (see e.g. Sambridge and Mosegaard 2002, Mosegaard and Tarantola 2002, Tarantola 2004). From a Bayesian point of view, geophysical model parameters are random variables whose distribution can be inferred by combining the prior density with the likelihood of observed data. In general, Bayesian estimation and inference have a number of advantages in statistical modelling and data analysis (Winkler 2003, Bolstad 2004, Gelman et al 2004; Congdon 2006; Robert 2007; Carlin and Louis 2008). The Bayes method formalises the process of learning from data, and provides confidence intervals on parameters and probability values on hypotheses. Using modern sampling methods such as Markov chain Monte Carlo, the Bayesian approach is readily adapted to complex random effects models that are more difficult to fit using classical methods (e.g. Carlin et al. 2001). The complete posterior distribution of the parameters resulting from Bayesian MCMC allows further analysis such as model selection and parameter uncertainty quantification. Increasingly, Bayesian MCMC finds new applications in geophysical inversion. Recent examples can be found in Mosegaard and Sambridge (2002), Tamminen (2004), Malinverno and Leaney (2005), Sambridge et al (2006), Fukuda and Johnson (2008), Chen et al (2008) and Bodin and Sambridge (2009).

In geophysical inverse problems we aim to reconstruct the distribution of subsurface properties (an 'earth model') given measurements that are usually acquired at the surface. These inverse problems are typically ill-posed, non-linear and non-unique. That is, significantly different earth models can fit equally well to the actual measurements. As clearly stated by Backus (1988), a geophysical inverse problem has an existence half, where we wish to obtain an earth model that fits the data, and a uniqueness half, where we want to measure how much the earth model may vary while fitting the data. A common way to solve the existence half of the inverse problem is to regularize, or damp the solution and obtain the smoothest earth model that fits the data within a given error. It is often necessary, however, to address the uniqueness half of the problem, and a way to quantify non-uniqueness or uncertainty is to use a probabilistic Bayesian inference approach. The starting point of Bayesian inference is to specify the initial uncertainty of the earth model parameters in a prior probability distribution. Bayes' rule then combines this prior distribution with a likelihood function (which measures how probable an earth model is in light of the measurements) to give a posterior distribution, which is the solution of the problem and describes the final uncertainty of the earth model (e.g. Tarantola & Valette 1982; Jackson & Matsu'ura 1985; Duijndam 1988, Mosegaard and Tarantola 1995; 2002, Tarantola 2004, Sambridge et al 2006, Bodin and Sambridge 2009).

There is a class of problems where the "number of unknowns is one of the unknowns". For these problems, a number of frameworks have been developed since the mid-1990s to extend the fixed-dimension MCMC to encompass trans-dimensional stochastic simulation. Among these trans-dimensional schemes, the reversible jump Markov chain sampling algorithm proposed by Green (1995) is certainly the most well understood and well developed. A survey of the state of the art on trans-dimensional Markov chain Monte Carlo can be found in Green (2003). A typical example of trans-dimensional problems in geophysics is the layered earth model, in which there are an unknown number of earth layers with each layer having unknown physical or geometrical properties such as mass density and layer thickness. The layered earth model has a close analogy to the change-point problem considered in the original paper on reversible jump Markov chain Monte Carlo (RJMCMC) by Green (1995). Typically, the change-point problem considers a number of step functions where neither the number of steps nor the locations of these steps are unknown, and associated with each step function there is an additional unknown parameter, the height.

Trans-dimensional MCMC has been successfully applied to the layered earth model by Malinverno (2002), where the parsimonious nature of Bayesian MCMC was demonstrated. That is, among all earth models that fit the data, those with fewer parameters (layers) have higher posterior probabilities. Sambridge et al (2006) applied RJMCMC as well as fixed dimension MCMC to both liner and nonlinear inverse problems, where it is demonstrated that by performing marginal likelihood calculations or evidence calculations, results of trans-dimensional sampling algorithms can be replicated with simpler fixed dimension MCMC sampling. The evidence calculations provide a useful diagnostic tool for investigating the number of degrees of freedom in the model inferred from measurement data. More recently, Bodin and Sambridge (2009) proposed a RJMCMC algorithm for a seismic tomography model parameterized using Voronoi cells. They have demonstrated that, among other findings, the uncertainty estimates based on posterior samples from RJMCMC can represent actual uncertainty very well.

In this paper we propose a RJMCMC sampling algorithm for a polygonal earth model, where the shape of a geological object is parameterized by a horizontal polygon, which is defined in the vertical plane and it is infinite in the direction perpendicular to the plane. The gravitational attraction of such an object can be measured from above the earth surface. The forward problem is computing the gravitational attraction on the surface, given the density contrast and the geometry of the structure. An inverse problem of practical interest is then to find the shape of the structure given measurements of gravitational attraction and the density contrast. A rapid computation algorithm for the forward problem was derived by Talwani et al (1959). The inverse problem was attempted by Chen et al (2006) using a hybrid-encoding genetic algorithm (HEGA). A key feature of HEGA is that the mutation operation is conducted in the decimal code and multi-point crossover operation in binary code. The main restrictions of applying HEGA to the polygonal earth model appear to be that the number of vertices is assumed to be known (fixed at the correct number), and also the search range for each vertex is confined locally in the neighbourhood of the true location of each vertex. It may be possible to design a special GA algorithm to deal with trans-dimensional problems. In the present work we remove both restrictions, i.e. the number of vertices is unknown, and the range of location for each vertex is not locally restricted. A fundamental difference between the algorithm presented here and that of Chen et al (2006) is that the former is a statistical inference and the later is a genetic optimization. Needless to say, both data uncertainty (measurement error) and model uncertainty, including the number of vertices, are naturally admitted in our Bayesian approach.

In Section 2, the general methodology and formulation of RJMCMC, as proposed by Green (1995), are described. Section 3 presents RJMCMC algorithm for the polygonal earth model, after introducing the parametric model and its forward problem. The various Markov chain proposal moves and the associated probability ratios and acceptance probabilities are discussed in detail. Section 4 applies the RJMCMC algorithm to the same problem as attempted by Chen et al (2006). Results, including model probabilities and posterior mode predictions on the shape of a gravity anomalous body are presented. Comparison between prior and posterior distributions of the number of vertices is made. A sensitivity study on prior distribution is also reported and discussed in Section 4. Section 5 concludes the paper.

## 2. Reversible jump Markov chain Monte Carlo

From a Bayesian point of view, there is no fundamental difference between observed data and parameters of a statistical model – all are considered random numbers from certain distributions. Given a *prior* distribution $\pi(\boldsymbol{\theta})$ and a likelihood $\pi(\mathbf{y}|\boldsymbol{\theta})$, the joint distribution of data $\mathbf{y}$ and the model parameter $\boldsymbol{\theta}$ can be expressed as

$$\pi(\mathbf{y},\boldsymbol{\theta}) = \pi(\mathbf{y}|\boldsymbol{\theta})\pi(\boldsymbol{\theta}).$$

Having observed data $\mathbf{y}$, the distribution of $\boldsymbol{\theta}$ conditional on $\mathbf{y}$, the *posterior* distribution, is determined by the Bayes theorem

$$\pi(\boldsymbol{\theta}|\mathbf{y}) = \frac{\pi(\mathbf{y}|\boldsymbol{\theta})\pi(\boldsymbol{\theta})}{\int \pi(\mathbf{y}|\boldsymbol{\theta})\pi(\boldsymbol{\theta})d\boldsymbol{\theta}}.$$

The posterior can then be used for predictive inference. Bayesian approach provides a consistent and robust mechanism for combining objective data observation and subjective judgment (e.g. geologists' expert opinions). The prior also allows the imposing of important geophysical, non-sample information such as the range (e.g. positivity, lower and upper bounds *etc*.) of certain parameters. There are a large number of useful texts on Bayesian inference, for example, see Robert and Smith (1994), Lee (1997), Berger (1999), Robert (2007), Winkler (2003), Gelman, Carlin, Stern and Rubin (2004), Bolstad (2004) and Carlin and Louis (2008).

The Markov chain Monte Carlo (MCMC) method provides a highly efficient alternative to traditional techniques by sampling from the posterior indirectly and performing the integration implicitly. The idea is straightforward: if the target distribution $\pi(.)$ is known only up to some multiplicative constant and it is sufficiently complex that we cannot sample from it directly, an indirect method of sampling from $\pi(.)$ is to construct an aperiodic and irreducible Markov chain whose stationary distribution is $\pi(.)$. Under certain regularity conditions, given in Roberts and Smith (1994) for example, the Markov chain mimics a random sample from $\pi(.)$. If the chain is sufficiently long, simulated values from the chain can be treated as a dependent sample from the target distribution and used to derive important summary statistics of $\pi(.)$.

A Markov process is said to show *detailed balance* if the transition rates between each pair of states $s$ and $s'$ in the state space obey

$$q(s,s')\pi(s) = q(s',s)\pi(s'),$$

where $q(s,s')$ is the Markov chain transition distribution or transition kernel, which is simply the conditional distribution of the next state $s'$ given the present state $s$. The detailed balance essentially means that the chain looks the same whether you run it forwards in time or backwards. A Markov process that satisfies the detailed balance equation is said to be a reversible Markov process or reversible Markov chain with respect to $\pi(.)$. This reversible property is desirable for any MCMC transition kernel to have, since any transition kernel for which the detailed balance holds will have stationary distribution $\pi(.)$.

*Reversible jump MCMC*
Under the framework outlined above, special considerations are required for problems where the number of unknowns is also one of the unknowns, i.e. we do not know how many parameters should be used to specify our model. Let $\{\mathcal{M}_k, k \in \mathcal{K}\}$ denote a countable collection of candidate models. Model $\mathcal{M}_k$ has a vector $\boldsymbol{\theta}_k$ of unknown parameters, $\boldsymbol{\theta}_k \in \mathcal{R}^{n_k}$, where the dimension $n_k$ may vary from model to model. Under a Bayesian framework, inference is carried out using the joint posterior distribution $\pi(\boldsymbol{\theta}_k, \mathcal{M}_k | \mathbf{y})$, where $\mathbf{y}$ is the observed data. The reversible jump Markov chain Monte Carlo (RJMCMC) proposed by Green (1995) provides a framework for constructing reversible Markov chain samplers that jump between parameter spaces of different dimensions, thus permitting exploration of joint parameter and model probability space via a single Markov chain.

As shown by Green (1995), detailed balance is satisfied if the proposed move from $(\boldsymbol{\theta}_i, \mathcal{M}_i)$ to $(\boldsymbol{\theta}_j, \mathcal{M}_j)$ is accepted with probability $\alpha = \min\{1, \alpha_{i \to j}(\boldsymbol{\theta}_i, \boldsymbol{\theta}_j)\}$, with $\alpha_{i \to j}(\boldsymbol{\theta}_i, \boldsymbol{\theta}_j)$ given by

$$\alpha_{i \to j}(\boldsymbol{\theta}_i, \boldsymbol{\theta}_j) = \frac{\pi(\boldsymbol{\theta}_j, \mathcal{M}_j | \mathbf{y}) r_{j \to i}(\boldsymbol{\theta}_j) q_{j \to i}(\boldsymbol{\theta}_j, \boldsymbol{\theta}_i)}{\pi(\boldsymbol{\theta}_i, \mathcal{M}_i | \mathbf{y}) r_{i \to j}(\boldsymbol{\theta}_i) q_{i \to j}(\boldsymbol{\theta}_i, \boldsymbol{\theta}_j)}, \qquad (1)$$

where $r_{i \to j}(\boldsymbol{\theta}_i)$ is the probability that a proposed jump from $\mathcal{M}_i$ to $\mathcal{M}_j$ is attempted, and $q_{i \to j}(\boldsymbol{\theta}_i, \boldsymbol{\theta}_j)$ is the density from which the proposed parameter $\boldsymbol{\theta}_j$ is drawn given $\boldsymbol{\theta}_i$

In general, it is difficult to find and specify a joint distribution $q_{i \to j}(\boldsymbol{\theta}_i, \boldsymbol{\theta}_j)$ directly to construct an efficient transition between models with different dimensions and interpretations of parameters. In practice the following two-step procedure is followed. In the first step a random vector $\mathbf{u}_i$ of dimension $m_i$ is drawn from a known density $\varphi_i(\mathbf{u} | \boldsymbol{\beta}_i)$ with parameter vector $\boldsymbol{\beta}_i$. Then the new state $\boldsymbol{\theta}_j$ is obtained from $\boldsymbol{\theta}_j = g_{i \to j}(\boldsymbol{\theta}_i, \mathbf{u}_i)$, where $g_{i \to j}(\boldsymbol{\theta}_i, \mathbf{u}_i)$ is a deterministic function of the current state $\boldsymbol{\theta}_i$ and the newly generated random vector $\mathbf{u}_i$. That is, $g_{i \to j}(\boldsymbol{\theta}_i, \mathbf{u}_i)$ is a pre-determined mapping function from $(\boldsymbol{\theta}_i, \mathbf{u}_i)$ to $\boldsymbol{\theta}_j$. The combination of a random generation of $\mathbf{u}_i$ and the deterministic mapping $\boldsymbol{\theta}_j = g_{i \to j}(\boldsymbol{\theta}_i, \mathbf{u}_i)$ provides a great flexibility for between model transitions.

The reverse move from $\mathcal{M}_j$ to $\mathcal{M}_i$ is constructed in a similar manner. That is, firstly random vector $\mathbf{u}_j$ of dimension $m_j$ is drawn from density $\varphi_j(\mathbf{u} | \boldsymbol{\beta}_j)$ with parameter vector $\boldsymbol{\beta}_j$, then $\boldsymbol{\theta}_i$ is proposed from $\boldsymbol{\theta}_i = g_{j \to i}(\boldsymbol{\theta}_j, \mathbf{u}_j)$, where $g_{j \to i}(\boldsymbol{\theta}_j, \mathbf{u}_j)$ is a deterministic function of the state $\boldsymbol{\theta}_j$ and random vector $\mathbf{u}_j$. The above described reversible moves satisfy the equal dimensionality condition $n_i + m_i = n_j + m_j$, i.e. the dimensions of $(\boldsymbol{\theta}_i, \mathbf{u}_i)$ and $(\boldsymbol{\theta}_j, \mathbf{u}_j)$ equal to each other. In addition, the mapping function is a diffeomorphism – both the transformation and its inverse are differentiable.

Following the two-step procedure as above, the ratio of the joint proposal densities in (1) can now be evaluated as

$$\frac{q_{j \to i}(\boldsymbol{\theta}_j, \boldsymbol{\theta}_i)}{q_{i \to j}(\boldsymbol{\theta}_i, \boldsymbol{\theta}_j)} = \frac{\varphi_j(\mathbf{u}_j | \boldsymbol{\beta}_j)}{\varphi_i(\mathbf{u}_i | \boldsymbol{\beta}_i)} \left| \frac{\partial g_{i \to j}(\boldsymbol{\theta}_i, \mathbf{u}_i)}{\partial (\boldsymbol{\theta}_i, \mathbf{u}_i)} \right|, \qquad (2)$$

where $\left|\partial g_{i \to j}(\boldsymbol{\theta}_i, \mathbf{u}_i) / \partial(\boldsymbol{\theta}_i, \mathbf{u}_i)\right|$ is the Jacobian of the deterministic mapping. Combining (1) and (2), the acceptance probability $\alpha_{i \to j}$ is then

$$\alpha_{i \to j} = \frac{\pi(\boldsymbol{\theta}_j, \mathcal{M}_j | \mathbf{y}) r_{j \to i}(\boldsymbol{\theta}_j) \varphi_j(\mathbf{u}_j | \boldsymbol{\beta}_j)}{\pi(\boldsymbol{\theta}_i, \mathcal{M}_i | \mathbf{y}) r_{i \to j}(\boldsymbol{\theta}_i) \varphi_i(\mathbf{u}_i | \boldsymbol{\beta}_i)} \left|\frac{\partial g_{i \to j}(\boldsymbol{\theta}_i, \mathbf{u}_i)}{\partial(\boldsymbol{\theta}_i, \mathbf{u}_i)}\right|. \quad (3)$$

Efficiency of RJMCMC depends on the choice of mapping function $g_{i \to j}$ and the proposal density $\varphi_i(.)$. In the last decade or so, much active research has been carried out to automate and optimize the between-model proposals for RJMCMC, see Green (2003), Brooks et al. (2003), Hastie (2004) and Fan et al. (2008). In general, careful design and implementation aimed at specific problems are required for successful applications of RJMCMC. One of the earliest applications of RJMCMC to geophysical inversion problems was due to Sambridge et al (2006). Figure 1 illustrates the main procedures of a typical RJMCMC simulation. For more accessible descriptions of RJMCMC under the context of geophysical inversion, see Malinverno (2002), Sambridge et al (2006) and Bodin and Sambridge (2009).

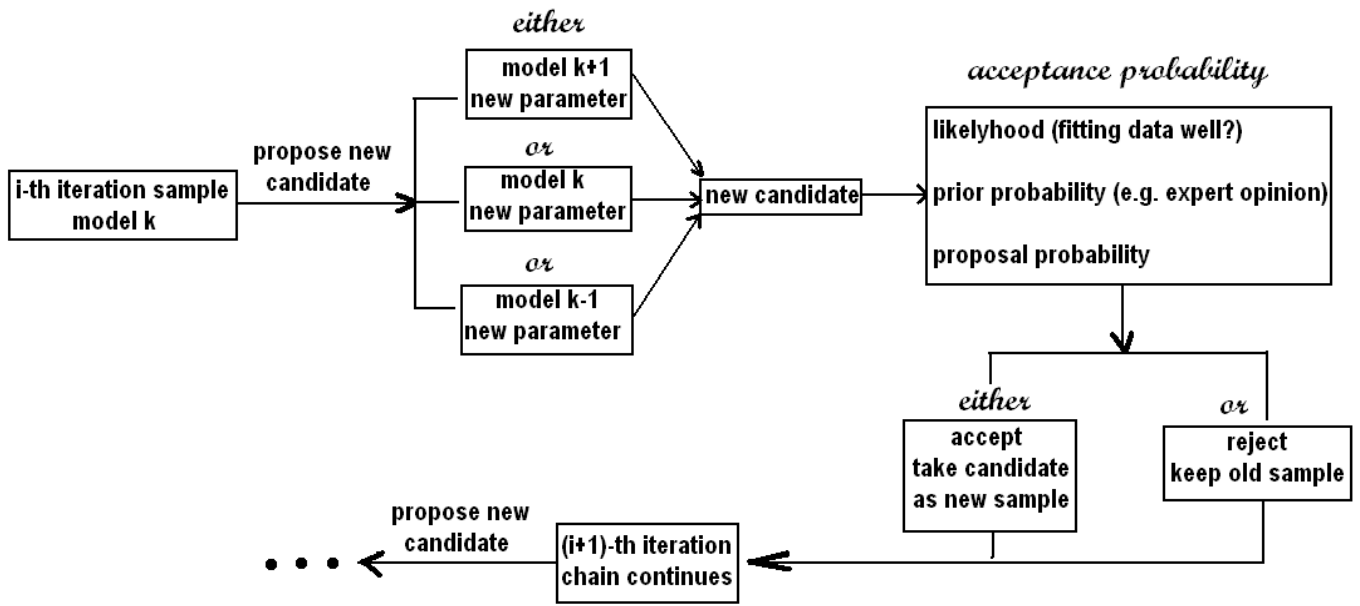

**Figure 1. Illustrative diagram of typical RJMCMC procedures.**

## 3. RJMCMC algorithm for a polygonal earth model

Many geological structures can be approximately analysed with a two-dimensional model. A two-dimensional horizontal column underneath the earth surface with a density contrast to its ambient is such an example. The gravitational attraction of such an object can be measured from above the earth surface. The forward problem is then to compute the gravitational attraction on the surface, given the density contrast and the geometry of the structure. The two-dimensional column is typically approximated by a polygon. Let $\boldsymbol{\theta}_k = \{\mathbf{v}_1, \mathbf{v}_2, ..., \mathbf{v}_k\}^T$ be the vector of coordinates for a *k*-sided polygon, identified by $\mathcal{M}_k$, with $\mathbf{v}_i = (x_i, z_i)$ and a clockwise ordering, as shown in Figure 2. For convenience we denote the vertex following $\mathbf{v}_i$ by $\mathbf{v}_i^+$ and the vertex before $\mathbf{v}_i$ by $\mathbf{v}_i^-$. Obviously each polygon model $\mathcal{M}_k$ has $2k$ parameters. Because the model identification number *k* for model $\mathcal{M}_k$ equals the number of vertices *k* for that model, the notation $\mathcal{M}_k$ is equivalent to *k* in the probability expressions in the remaining texts.

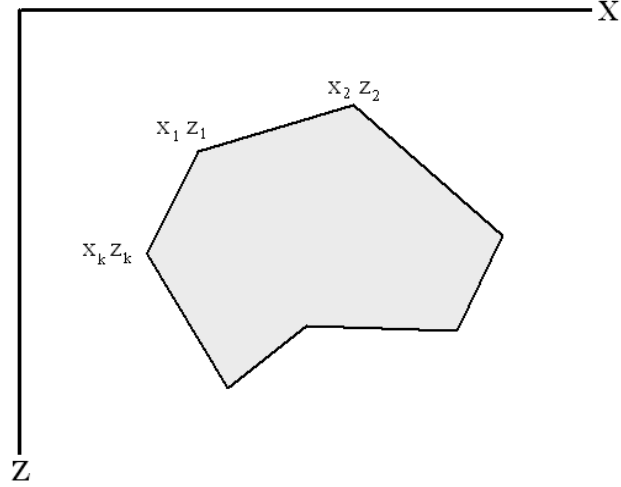

**Figure 2. Illustration of the geological model of a horizontal polygonal column.**

## 3.1. The forward problem

The forward problem has been solved analytically by Talwani et al. (1959), where from the integral expression for the gravitational attraction of a two-dimensional object, due to Hubbert (1948), an analytical expression is derived for an arbitrary polygon. The following alternative expression for the gravitational attraction $\Delta g(\tilde{x}, \tilde{z})$ at $(\tilde{x}, \tilde{z})$ is given by Chen et al (2006)

$$\Delta g(\tilde{x}, \tilde{z}) = 2G\Delta\rho \sum_{i=1}^{k} \frac{\xi_i \eta_{i+1} - \xi_{i+1} \eta_i}{(\xi_{i+1} - \xi_i)^2 + (\eta_{i+1} - \eta_i)^2} \times \left[ \frac{1}{2}(\eta_{i+1} - \eta_i) \ln \frac{\xi_{i+1}^2 + \eta_{i+1}^2}{\xi_i^2 + \eta_i^2} + g_i \right], \quad (4)$$

$$g_i = (\xi_{i+1} - \xi_i)(\tan^{-1} \frac{\xi_{i+1}}{\eta_{i+1}} - \tan^{-1} \frac{\xi_i}{\eta_i}), \quad \xi_i = x_i - \tilde{x}, \quad \eta_i = z_i - \tilde{z}$$

where *G* is gravitational constant, $\Delta\rho$ is the density contrast of the polygon. Note we have to define $x_{k+1} = x_1$ and $z_{k+1} = z_1$ to make (4) correct at $i = k$, i.e. the last line segment is formed by vertices $(x_k, z_k)$ and $(x_1, z_1)$.

Typically the measurement of $\Delta g$ is done on the surface at $\tilde{z} = 0$ between a certain range $\tilde{x}_{\min} \leq \tilde{x} \leq \tilde{x}_{\max}$. The inverse problem is then, given discrete measurements of $\Delta g$ at $m$ locations $\tilde{x}_1, \tilde{x}_2, ..., \tilde{x}_m$, detect the polygon shape underneath the surface, i.e. estimate the polygon model $\boldsymbol{\theta}_k$ or $(\mathbf{v}_1, \mathbf{v}_2, ..., \mathbf{v}_k)$. Significantly, we do not know the number of vertices $k$ of the polygon, which naturally is one of the unknowns. Broadly speaking, this inverse problem can be formulated as a trans-dimensional model selection problem, for which the RJMCMC is especially suitable.

The acceptance probability (1) is factored into posterior ratio and proposal ratio. Given a prior $\pi(\boldsymbol{\theta}_k, \mathcal{M}_k)$, a likelihood $\pi(\mathbf{y} | \boldsymbol{\theta}_k, \mathcal{M}_k)$ and the observed data $\mathbf{y}$, the joint posterior distribution of model $\mathcal{M}_k$ and its parameter $\boldsymbol{\theta}_k$ conditional on $\mathbf{y}$ is determined by the Bayes theorem

$$\pi(\boldsymbol{\theta}_k, \mathcal{M}_k | \mathbf{y}) = \frac{\pi(\mathbf{y} | \boldsymbol{\theta}_k, \mathcal{M}_k) \pi(\boldsymbol{\theta}_k, \mathcal{M}_k)}{\sum_{i=1}^{M} \pi(\mathcal{M}_i) \int \pi(\mathbf{y} | \boldsymbol{\theta}_i, \mathcal{M}_i) \pi(\boldsymbol{\theta}_i) d\boldsymbol{\theta}_i}, \tag{5}$$

where $M$ is the number of competing models. Now consider the transition from model $\mathcal{M}_i$ to model $\mathcal{M}_j$, from (5) the posterior ratio can be further factored into prior ratio and likelihood ratio

$$\frac{\pi(\boldsymbol{\theta}_j, \mathcal{M}_j | \mathbf{y})}{\pi(\boldsymbol{\theta}_i, \mathcal{M}_i | \mathbf{y})} = \frac{\pi(\boldsymbol{\theta}_j, \mathcal{M}_j)}{\pi(\boldsymbol{\theta}_i, \mathcal{M}_i)} \cdot \frac{\pi(\mathbf{y} | \boldsymbol{\theta}_j, \mathcal{M}_j)}{\pi(\mathbf{y} | \boldsymbol{\theta}_i, \mathcal{M}_i)}. \tag{6}$$

Thus the acceptance probability $\alpha_{ij}$ is factored as three ratios – the prior ratio, the likelihood ratio and the proposal ratio. In the following sections we discuss each of the three components.

### 3.2. Likelihood function

Assuming a Gaussian noise with standard deviation $\sigma$ in the measured data $y_j = \Delta g_j, j = 1,...,m$, the likelihood for a given model $\mathcal{M}_k$ is given by

$$\pi(\mathbf{y} | \boldsymbol{\theta}_k, \mathcal{M}_k) = \exp\left(-\frac{\sum_{j=1}^{m}(y_j - \hat{y}(\boldsymbol{\theta}_k, \tilde{x}_j))^2}{2\sigma^2}\right), \tag{7}$$

where $\hat{y}(\boldsymbol{\theta}_k, \tilde{x}_j) = \Delta g(\tilde{x}_j, 0)$ is the forward prediction by model $\mathcal{M}_k$, given in (4), and $y_j$ is observation or measurement of gravitational attraction on the surface at location $\tilde{x}_j$. If the noises are assumed to be correlated, $\sigma$ in (7) is replaced with the assumed covariance matrix.

### 3.3. Prior density

Similar problems dealing with geometrical shapes have been studied under image analysis and pattern recognition contexts, e.g. Grenander and Miller (1994), Qian et al (1996) and Pievatolo and Green (1998), just to name a few. Building a proper statistical model of polygons is a non-trivial task, mainly due to difficulties associated with constraints on the line segments (non-self-intersecting, for instance), consequently a measure space for all polygons is not easily obtained. A proper prior distribution for polygons has to be based on a proper definition of a measure space for polygons. For detailed discussions see Arak et al (1993) and Pievatolo and Green (1998). Here we adapt the prior model similar to that proposed by Pievatolo and Green (1998)

$$\pi(\pmb{\theta}_k, \mathcal{M}_k) \propto \exp\left(-k^\gamma - \frac{1}{k}\sum_{i=1}^k [\phi_i(\pmb{\theta}_k) - \omega_k]^2\right), \quad \omega_k = (k-2)\pi/k, \quad k \geq 3, \quad (8)$$

where $\gamma$ satisfies $\gamma \geq 1$ in order to have a proper prior, $\phi_i(\pmb{\theta}_k)$ is the angle in radians interior to the $i$th vertex of polygon $\pmb{\theta}_k$. The prior given in (8) is a joint prior of $k$ and the shape of the polygon with $k$ vertices. As a shape density, (8) is invariant with respect to the position, the orientation and the length scale of the polygon. The term containing the angles penalizes irregular polygons, as $\omega_k$ is the constant interior angle of a regular $k$-sided polygon. In other words, for any $k \geq 3$, the regular polygon maximizes the shape density (8). As $k \to \infty$, it is evident the shape maximizing (8) is a circle. In Pievatolo and Green (1998), $\omega_k$ was set to constant $\pi$ and two more hyperparameters were used in addition to $\gamma$. The term $k^\gamma$ penalizes an increasing number of vertices, assigning a higher probability to parsimonious polygons. An increasing value of $\gamma$ favours simpler polygons. Pievatolo and Green (1998) gave a detailed discussion on the sufficient conditions on $\gamma$ for the prior distribution to be proper.

From geophysical point of view, the prior density (8) may not be ideal if it strongly favours regular polygons, since we know *a priori* the shape of a geophysical structure in nature is more likely to be irregular. On the other hand, without penalizing any irregularity in the angles, a significant proportion of polygon samples drawn from the Markov chain sampling could have acute angles close to zero degrees or obtuse angles close to 180 degrees, or worse they are inadmissible due to self tangling.

As a shape density, (8) does not impose any constraints on the length scale and location of the polygon. We assume the polygon is contained in a rectangle $(0 \leq x \leq x_{max}, 0 \leq z \leq z_{max})$, and it is reasonable to let $x_{max} = \tilde{x}_{max}$, the last measurement location. Obviously if the object extends beyond $\tilde{x}_{max}$, the measurement coverage is insufficient. The maximum depth $z_{max}$ can be determined by expert opinion. If the solution for the polygon is close to the boundaries $x = x_{max}$ and $z = z_{max}$, it is an indication the bounds may not be large enough to well contain the solution.

### 3.4. Markov chain moves

It is difficult, if not impossible, to make truly random moves of vertices which always result in admissible polygons. The challenge is to devise moves enabling efficient mixing of Markov chains under the constraints of admissible polygons, e.g. no self intersecting of polygon sides. Fortunately, the requirement of maintaining a reasonable acceptance rate of Markov chain samples among proper polygons is consistent with the requirement of low rate of rejection of illegal polygons.

Instead of imposing explicitly the constraints to always yield proper polygons, in this study a straightforward rejection sampling approach is adopted. As described below, the step size of the random walk in both the within-model move and the between-model move is restricted by a variance tied to the length of the intersecting sides of the polygon. This is to ensure a high rate of success in generating admissible (no self tangling) polygons. Illegal polygons resulting from any reversible jump moves are simply rejected. Numerical experiments show our sampling strategy yields about 80% admissible polygons. The following three types of moves are proposed for the Markov chain:

1. Within-model move - shift a vertex to a nearby location.
2. Birth move - for all $k < k_{max}$, split a randomly chosen side into two at the middle, and make a type 1 move from the middle point.
3. Death move - for all $k > 3$, delete a randomly chosen vertex and form a new side by joining the two neighbouring vertices.

Move types 1 is a regular within-model Metropolis-Hastings move (Metropolis et al 1953, Hastings 1970). Move types 2 and 3 are trans-dimensional moves and require a different method such as the reversible jump algorithm proposed by Green (1995). Below we describe details of constructing each of the RJMCMC moves specifically for our problem.

### 3.4.1. Within-model proposal

Move type 1 changes the location of a randomly chosen vertex $\mathbf{v}_i$ from $(x_i, z_i)$ to $(x'_i, z'_i)$. The new location is proposed by drawing two independent random numbers $r$ and $\vartheta$ – one for distance and one for direction. The random distance $r$ is drawn from a normal distribution $F_N(0, \sigma_a)$ with zero mean and variance $\sigma_a = \min(d_i^-, d_i^+)C_a$, where $d_i^-$ and $d_i^+$ are the lengths of the two polygon lines intersecting at vertex $\mathbf{v}_i$ and $C_a$ is a constant. The random angle $\vartheta$ is drawn from uniform $(0, 2\pi)$. Then the new position $\mathbf{v}'_i(x'_i, z')$ is determined as

$$x'_i = x_i + r\cos(\vartheta), \quad z'_i = z_i + r\sin(\vartheta). \tag{9}$$

Instead of left-truncating the normal distribution $F_N(0, \sigma_a)$ at zero to keep $r$ positive, we let $r$ be drawn from the entire support, so it can take negative values. From (9), a negative value of $r$ is equivalent to taking the absolute value of $r$ and adding to (or subtracting from) $\vartheta$ a constant angle of $\pi$. This is more efficient than left-truncating the normal distribution. The above proposal attempts to keep the new location randomly scattered around the existing location whilst also reducing the possibility of tangling of non-neighbouring lines in the polygon. Figure 3 illustrates the within-model move. As shown in Figure 3, with $C_a = 0.25$ the variance $\sigma_a = \min(d_i^-, d_i^+)C_a$ ensures that with a probability greater than 0.95 (corresponding to two standard deviation), the new position will be inside the shaded circle.

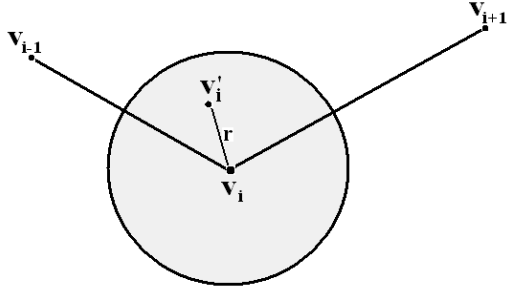

**Figure 3. Illustration of the type 1 move.**

The proposal density is

$$q(\boldsymbol{\theta}'_k | \boldsymbol{\theta}_k) = f_n(r|0,\sigma_a) \cdot (1/2\pi) \cdot r,$$

where $f_n(r|0,\sigma_a)$ is the normal density, $(1/2\pi)$ is the uniform density for the angle $\vartheta$ and $r = \partial(x'_i, z'_i)/\partial(r,\vartheta)$ is the Jacobian for the density transformation. The proposal density ratio $q(\boldsymbol{\theta}_k | \boldsymbol{\theta}'_k)/q(\boldsymbol{\theta}'_k | \boldsymbol{\theta}_k)$ is simply the ratio $f_n(r|0,\sigma'_a)/f_n(r|0,\sigma_a)$ for the within-model move, where $\sigma'_a$ is the variance defined the in same way as $\sigma_a$, but calculated with the lengths of the two polygon lines intersecting at the new vertex position $\mathbf{v}'_i$.

### 3.4.2. Birth-death proposals

From (3), the birth move has the following proposal ratio

$$\frac{d_{k+1}\varphi_{k+1}(\mathbf{u}_{k+1}|\boldsymbol{\beta}_{k+1})}{b_k\varphi_k(\mathbf{u}_k|\boldsymbol{\beta}_k)} \left|\frac{\partial g_{k\to k+1}(\boldsymbol{\theta}_k,\mathbf{u}_k)}{\partial(\boldsymbol{\theta}_k,\mathbf{u}_k)}\right|, \tag{10}$$

where $d_{k+1} = r_{k+1 \to k}$ is the probability of attempting the death move from $\mathcal{M}_{k+1}$ to $\mathcal{M}_k$, and $b_k = r_{k \to k+1}$ is the probability of attempting the birth move from $\mathcal{M}_k$ to $\mathcal{M}_{k+1}$. For the birth move, we first randomly chose a side connecting vertices $\mathbf{v}_i$ and its next neighbour $\mathbf{v}_i^+$, and split the side at the middle. A type 1 move is then made for this middle point to complete the birth move. The vector of the auxiliary variables for the birth move is $\mathbf{u}_k = (r,\vartheta)$. The parameter vector associated with these auxiliary variables is $\boldsymbol{\beta}_k = (\sigma_a)$. Thus the proposal density for the birth move can be expressed as

$$\varphi_k(\mathbf{u}_k,\boldsymbol{\beta}_k) = \varphi_k(r,\vartheta | \sigma_a) = (1/k)f_n(r|0,\sigma_a)(1/2\pi). \tag{11}$$

Denote the new point generated in the birth-move by $\mathbf{v}^* = (x^*, z^*)$, the temporary parameter state for $\mathcal{M}_{k+1}$ is then $\boldsymbol{\theta}^*_{k+1} = g_{k \to k+1}(\boldsymbol{\theta}'_k, \mathbf{u}_k) = (x_1, z_2, ..., x_i, z_i, x^*, z^*, x_{i+1}, z_{i+1}, ..., x_k, z_k)^T$. The relationship between $\mathbf{v}^*$ and $\mathbf{v}_i$ and $\mathbf{v}_i^+$ is, from the split and type 1 move combination

$$x^* = 0.5(x_i^- + x_i^+) + r\cos(\vartheta), \quad z^* = 0.5(z_i^- + z_i^+) + r\sin(\vartheta), \tag{12}$$

from which the Jacobian in (10) is found to be a very simple expression

$$\left| \frac{\partial g_{k \to k+1}(\boldsymbol{\theta}_k, \mathbf{u}_k)}{\partial(\boldsymbol{\theta}_k, \mathbf{u}_k)} \right| = r. \tag{13}$$

For the death move from $\mathcal{M}_{k+1}$ to $\mathcal{M}_k$, a randomly chosen vertex $\mathbf{v}_i, 1 \leq i \leq k+1$ is removed, and the neighbouring vertices $\mathbf{v}_i^{-1}$ and $\mathbf{v}_i^+$ are connected to form a new side. In this case the density $\varphi_{k+1}(\mathbf{u}_{k+1} | \boldsymbol{\beta}_{k+1}) = 1/(k+1)$. Finally the proposal ratio (10) for the birth-death move is

$$\frac{2\pi d_{k+1} k r}{b_k(k+1) f_n(r | 0, \sigma_a)}. \tag{14}$$

For the death-birth move, the ratio is inversed from (14), and in this case vertex $\mathbf{v}_i$ is removed the radius $r$ is the distance from vertex $\mathbf{v}_i$ to the middle point between $\mathbf{v}_i^-$ and $\mathbf{v}_i^+$.

### 3.5. Acceptance probability

Combining the various ratios described in Sections 3.2, 3.3 and 3.4, the acceptance probabilities for the various moves are summarised as follows

***Within-model moves***

$$\alpha = \min\left\{1, \frac{\pi(\boldsymbol{\theta}_j, \mathcal{M}_j)}{\pi(\boldsymbol{\theta}_i, \mathcal{M}_i)} \frac{\pi(\mathbf{y} | \boldsymbol{\theta}_k', \mathcal{M}_k)}{\pi(\mathbf{y} | \boldsymbol{\theta}_k, \mathcal{M}_k)}\right\}, \tag{15}$$

***Birth moves***
Corresponding to the birth proposal, the acceptance probability is

$$\alpha = \min\left\{1, \frac{\pi(\boldsymbol{\theta}_{k+1}, \mathcal{M}_{k+1})}{\pi(\boldsymbol{\theta}_k, \mathcal{M}_k)} \frac{\pi(\mathbf{y} | \boldsymbol{\theta}_{k+1}, \mathcal{M}_{k+1})}{\pi(\mathbf{y} | \boldsymbol{\theta}_k, \mathcal{M}_k)} \frac{2\pi d_{k+1} k r}{b_k(k+1) f_n(r | 0, \sigma_a)}\right\}. \tag{16}$$

***Death moves***
Paring with (16) the acceptance probability for the death move is

$$\alpha = \min\left\{1, \frac{\pi(\boldsymbol{\theta}_k, \mathcal{M}_k)}{\pi(\boldsymbol{\theta}_{k+1}, \mathcal{M}_{k+1})} \frac{\pi(\mathbf{y} | \boldsymbol{\theta}_k, \mathcal{M}_k)}{\pi(\mathbf{y} | \boldsymbol{\theta}_{k+1}, \mathcal{M}_{k+1})} \frac{b_k(k+1) f_n(r | 0, \sigma_a)}{2\pi d_{k+1} k r}\right\}. \tag{17}$$

In (15), (16) and (17) the likelihood $\pi(\mathbf{y}|\boldsymbol{\theta}_k, \mathcal{M}_k)$ is given by (7) and the prior density $\pi(\boldsymbol{\theta}_j, \mathcal{M}_j)$ by (8). For each Markov chain step, an independent random choice is made among the three possible moves. Obviously the probability of within-model move is $1-(b_k+d_k)$. Naturally at the boundaries we have $d_3 = b_M = 0$, i.e. there is no death move at the minimum dimension and no birth move at the maximum dimension.

## 4. Numerical experiments

As an application of the RJMCMC algorithm we solve the same problem attempted by Chen et al (2006), where they used a hybrid-encoding genetic algorithm. Figure 4 shows, among other things, the true shape of the object, which is a ten-sided polygon and is labelled Model-3 in Chen et al (2006). Exact data of the coordinates for the true model are given in Table 1. It is worth pointing out the three vertices numbered 4, 5 and 6 in Table 1 are nearly on the same line, so the true model appears to be a 9-sided polygon.

The main difference between the present study and the work of Chen et al (2006), apart from the totally different approach, is that here we assume the number of vertices is unknown, thus the problem is trans-dimensional and we do not have any prior knowledge of whereabouts of any of the vertices, except a wide global range which could be readily deduced in practice from the location of the measured gravity anomaly. Without loss of generality we present the forward problem as dimensionless, i.e. we assume the dimensionless quantity $\Delta g^* = \Delta g(\tilde{x}, \tilde{z})/2G\Delta\rho$ is measured with a certain standard error $\sigma^*$. We refer $\Delta g^*$ as the gravity anomaly. We further assume 21 measurement points are uniformly located on the surface in the range $0 \leq \tilde{x}_j \leq 500$, i.e. $\tilde{x}_j = 25(j-1)$, $\tilde{z}_j = 0$, $j=1,...,21$. In all calculations $\sigma^*$ is fixed at $0.2$, which is just under 10% of the minimum value of $\Delta g^*$ measured at the last point $\tilde{x}_j = 500$.

**Table 1. Geometry data of the true polygonal model.**

| Geometry of true model | | | | | | | | | | |
|---|---|---|---|---|---|---|---|---|---|---|
| vertex | 1 | 2 | 3 | 4 | 5 | 6 | 7 | 8 | 9 | 10 |
| $x(m)$ | 110 | 220 | 290 | 390 | 310 | 250 | 220 | 180 | 170 | 150 |
| $z(m)$ | 20 | 10 | 20 | 20 | 30 | 40 | 60 | 60 | 40 | 40 |

### 4.1. MCMC simulation details

For model dimension $k$ (the number of vertices of the polygon) and the shape of a polygon, their joint prior is as discussed in Section 3.3. We first set $\gamma = 1.6$, the same as that used in Pievatolo and Green (1998) in a different setting. Later on different $\gamma$ values are used for a sensitivity study. The only other prior knowledge about the model is that we impose a global rectangular boundary for the location of the polygon. That is, the polygon is contained within the rectangle $(0 \leq x \leq x_{max} = 500, 0 \leq z \leq z_{max} = 100)$. Note in practice this boundary can usually be correctly estimated from the measured anomaly and experience. The key requirement for imposing such a boundary is that it should be large enough to contain the true model.

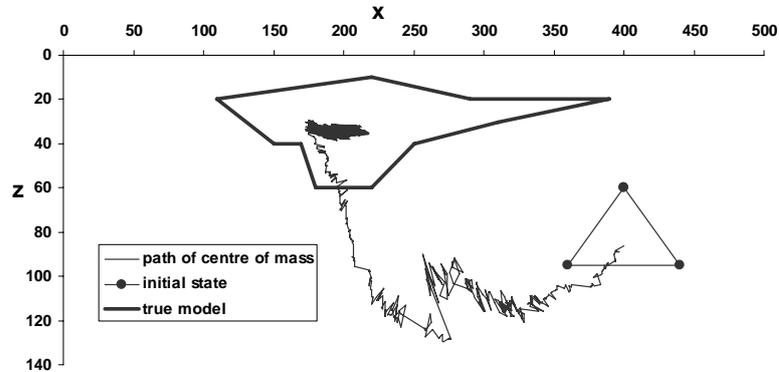

**Figure 4. The true model, the initial Markov chain state and the sample path of the centre of mass of the polygon vertices during the burn-in period.**

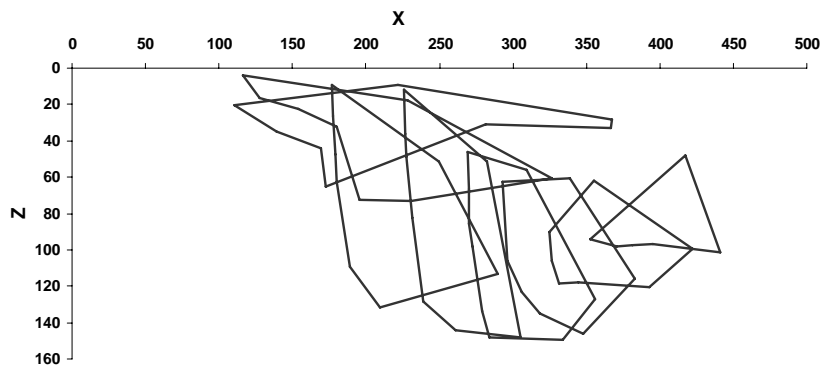

**Figure 5. A sequence of polygons with varying dimensions during the burn-in period – the adjacent samples are spaced at least 2000 steps apart.**

For the initial state of the chain we assume an arbitrary triangle, the simplest polygon, as shown in Figure 4. We deliberately choose a starting polygon with its location, shape and length scale very different from the true model to test the robustness of the numerical algorithm. All RJMCMC simulations started from the same initial state represented by the triangle shown in Figure 4. In real applications, we could choose a more proper starting polygon based on the anomaly data. For example, we could locate the polygon directly below the peak of the anomaly curve, and estimate a length scale from the horizontal spread of the gravity anomaly curve (e.g. the distance from the left half peak point to the right half peak point).

The burn-in iterations are set at 100,000, that is, the first 100,000 samples from the chain are discarded. The adequacy of the burn-in length can be confirmed by monitoring the path of the centre of mass of the polygon vertices – the path should become localized in the later stage of the burn-in period. Figure 4 also shows the sample path for the centre of mass during the burn-in period. It is evident the centre of mass is indeed confined to a local region in the later stage of the burn-in period, as it should be if the stationary posterior distribution region is reached by the chain. Figure 5 shows a sequence of RJMCMC polygon samples with varying number of dimensions during the burn-in period.

After the burn-in, 1,000,000 steps were run for the Markov chain. Typically, a single long chain is adequate to explore the posterior density, provided allowance is made for dependence in the samples (e.g. Gilks et al 1996). For simpler MCMC problems of fixed dimension, usually the number of simulations in the order of 100,000 is sufficient. For the current RJMCMC problem, however, much longer chain is required for adequate convergence. The time taken for 100,000 RJMCMC iterations is about 8.5 minutes on a desktop PC with an Inter.Xeron CPU at 2.93GHz, so it could take up to 2 hours for running a chain of one million steps. For even more computing intensive three-dimensional or multiple objects problems, a fast solver for the forward problem, a good choice of prior function and an efficient parameterization are all required to ensure that RJMCMC can obtain results within a reasonable time frame.

### *4.2. Model probability*

Figure 6 shows the model probabilities predicted by posterior samples from RJMCMC (the shaded bars). The model with the highest probability is the 8-sided polygon, model $\mathcal{M}_8$, closely followed by the 9-sided, then the 7-sided and the 10-sided models in decreasing probability order. The four models with the highest probabilities altogether sum to 99% of total probability. That is, it is almost surely that the polygon dimension is in the range $7 \leq k \leq 10$, according to the posterior distribution obtained from RJMCMC simulation. This compares well with the actual dimension of the true model, which is almost a 9-sided polygon. Note that in Figure 6 some of the probabilities are so low, e.g. $\Pr(k = 4) < 0.001$, they are not visible in the graph.

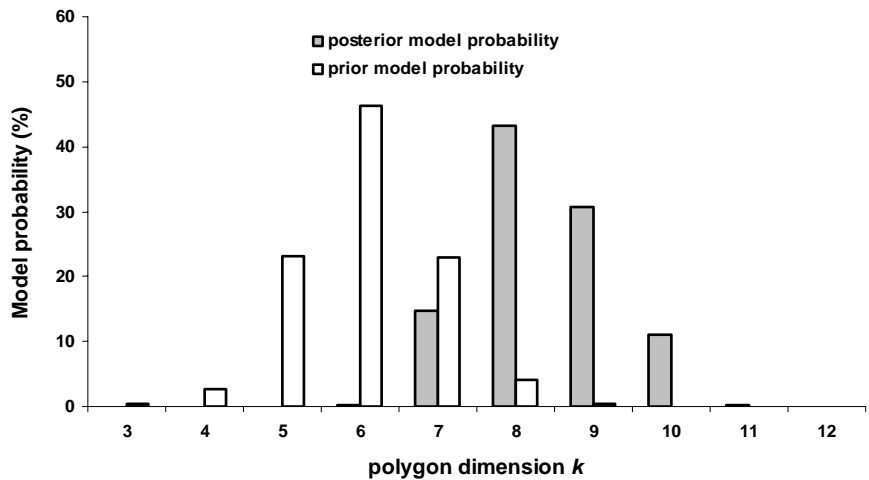

**Figure 6. Model probabilities predicted by MCMC posterior distribution and prior distribution.** $\gamma = 1.6$.

As an interesting comparison, Figure 6 also shows the model probability predicted by prior distribution alone (unshaded bars). In this case the same RJMCMC simulation was carried out with likelihood ratio fixed at unity, i.e. no data influence or no geophysical model was present. The prior predicted a 46% probability for the six-sided polygon, while the posterior probability for the six-sided polygon is virtually zero (less than 1%). Model $\mathcal{M}_8$, the 8-sided polygon has a posterior probability of 43% , but it has a prior probability of only 4% . These significant differences between the posterior prediction and the prior prediction are obviously caused by data. The comparison clearly shows that the influence of data shifts the model probability much closer to the true model.

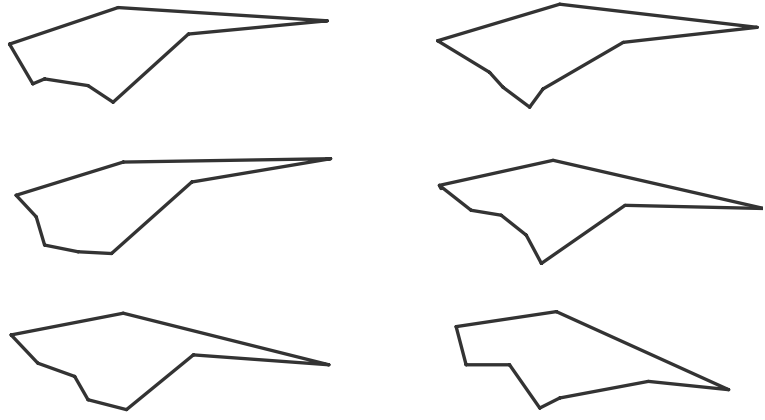

**Figure 7. Some posterior samples for model $\mathcal{M}_8$.**

## *4.3. Posterior estimations*

The MCMC solution provides samples from the posterior distribution for each model $\mathcal{M}_k$. Figure 7 shows some arbitrarily selected samples for model $\mathcal{M}_8$. The Bayesian estimators of particular interest from MCMC are the maximum *a posterior* (MAP) estimator and the minimum mean square error (MMSE) estimator, defined as follows

$$\text{MAP}: \quad \hat{\boldsymbol{\theta}}^{MAP} = \arg\max_{\boldsymbol{\theta}}[\pi(\boldsymbol{\theta}\,|\,\mathbf{y},\mathcal{M}_k)], \tag{18}$$

$$\text{MMSE}: \quad \hat{\boldsymbol{\theta}}^{MMSE} = E[\boldsymbol{\theta}\,|\,\mathbf{y},\mathcal{M}_k]. \tag{19}$$

The MAP and MMSE estimators are the posterior mode and mean respectively in a Bayesian setting. From the perspective of solving an inverse problem or optimization problem, we can take the posterior mode or mean as our solution. The posterior mode for model $\mathcal{M}_8$ and $\mathcal{M}_9$ is shown in Figure 8 and 9, respectively. The corresponding forward predictions of $\mathcal{M}_8$ and $\mathcal{M}_9$ are compared with that of the true model (exact measurement without error) in Figure 10. The very poor forward prediction by the starting triangle (shown in Figure 4) is also shown in Figure 10. From these results we make the following observations and comments:

- The estimated polygon model by the posterior mode gives, on the whole, a reasonably close presentation of the true shape of the object, although there is still noticeable difference between the true model and the predictions in some of the details of the object;

- $\mathcal{M}_8$ and $\mathcal{M}_9$ (and $\mathcal{M}_{10}$, picture not shown) are similar overall. Compared with $\mathcal{M}_8$, models with higher dimensions or more parameters do not necessarily give a better presentation of the true model. Details not resolved by the simpler model are also not resolved by the more complicated models. Thus RJMCMC appears to be parsimonious, in the sense that, for a given prior, among the models that satisfactorily represent the true model, the model with the higher posterior probability has less number of parameters. This kind of seemingly "natural" parsimony property of Bayesian MCMC approach was also demonstrated in a layered earth model problem (Malinverno 2002). It is worth pointing out that parsimony of Bayesian models is typically enforced through by a proper prior;
- Forward predictions of both $\mathcal{M}_8$ and $\mathcal{M}_9$ agree very well with that of the true model, despite the two differ from each other and both models have some noticeable difference with the true model. This is also true among other models with lower or higher dimensions. This demonstrates the ill-conditioning of the inverse problem in consideration – multiple inverse solutions can well satisfy the same forward conditions.

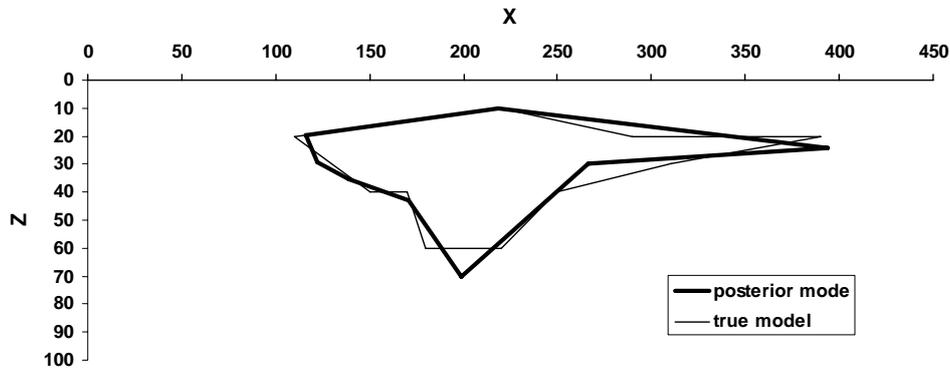

**Figure 8. Posterior mode for model $\mathcal{M}_8$, $\gamma = 1.6$.**

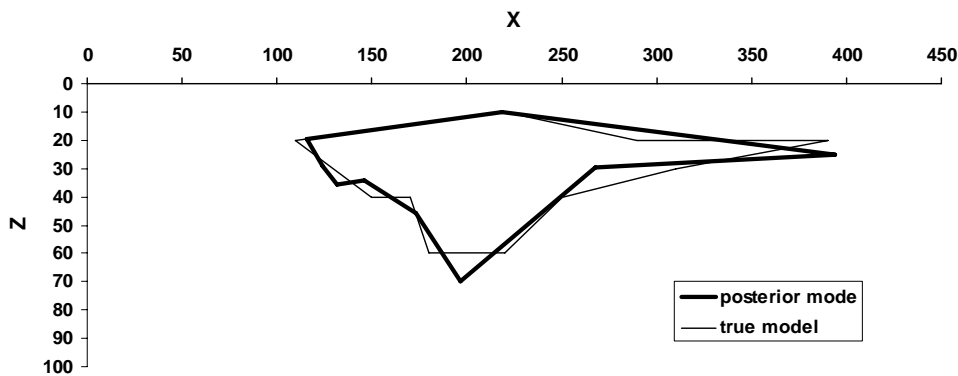

**Figure 9. Posterior mode for model $\mathcal{M}_9$, $\gamma = 1.6$.**

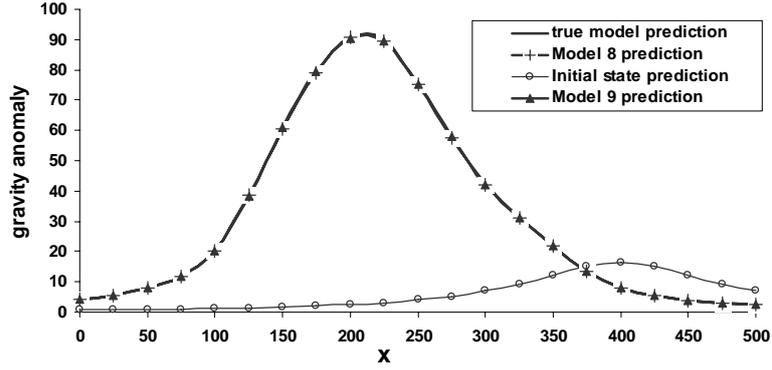

**Figure 10. Comparison of forward predictions of gravity anomaly.**

The model uncertainty is quantified by the model probability as shown in Figure 6. There are various ways to quantify parameter uncertainty conditional on a given model, for example see Bodin and Sambridge (2009). One of the straightforward quantifications is using the coefficient of variance (CV) of the posterior samples, which is defined as the ratio of standard deviation over the mean. For model $\mathcal{M}_8$, the most likely model, the maximum CV is found to be 0.37, occurring in one of the vertical coordinates. The minimum CV is 0.04 corresponding to a horizontal coordinate. The average CV for vertical coordinates is 0.29, while the average CV for horizontal coordinates is 0.09. For models $\mathcal{M}_7$, $\mathcal{M}_9$ and $\mathcal{M}_{10}$ the CV values are very similar to those of model $\mathcal{M}_8$. The significantly higher CV for the vertical coordinates reflects that the forward prediction of gravity anomaly is less sensitive to the depth of vertices than to the horizontal positions, at least for the case when the gravity anomaly measurement points are horizontally located.

## *4.4. Sensitivity of prior*

Results so far are for parameter $\gamma = 1.6$ with the prior given by (8). A sensitivity study was carried out to see the effect of different $\gamma$ values. The same RJMCMC simulations with $\gamma = 1.2$ (a 25% decrease in $\gamma$) and $\gamma = 2.0$ (a 25% increase in $\gamma$) were run. Results of model probability for $\gamma = 1.2$ are shown in Figure 11. Comparing Figure 11 with Figure 6, there is a slight upward shift of model probability towards polygons with higher number of vertices. This shift occurs in both prior and posterior predictions of model probability, and it is consistent with our expectation that a smaller value of $\gamma$ imposes less restrictions on the number of vertices $k$. The posterior mode prediction of the most likely model $\mathcal{M}_9$ for $\gamma = 1.2$ is shown in Figure 12, from which we see the overall agreement with the true shape is still satisfactory, but the agreement is not as good as $\mathcal{M}_9$ for $\gamma = 1.6$ shown in Figure 9.

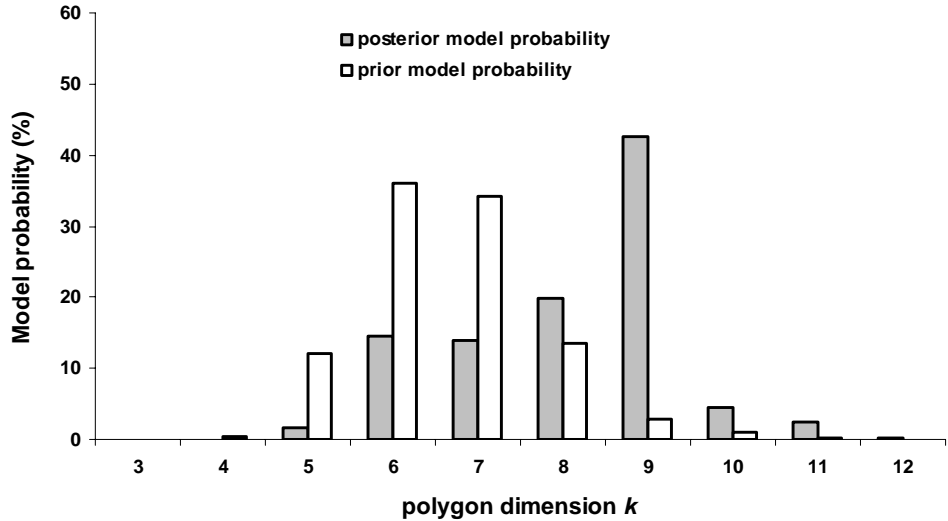

**Figure 11. Model probabilities predicted by MCMC posterior distribution and prior distribution.** $\gamma = 1.2$.

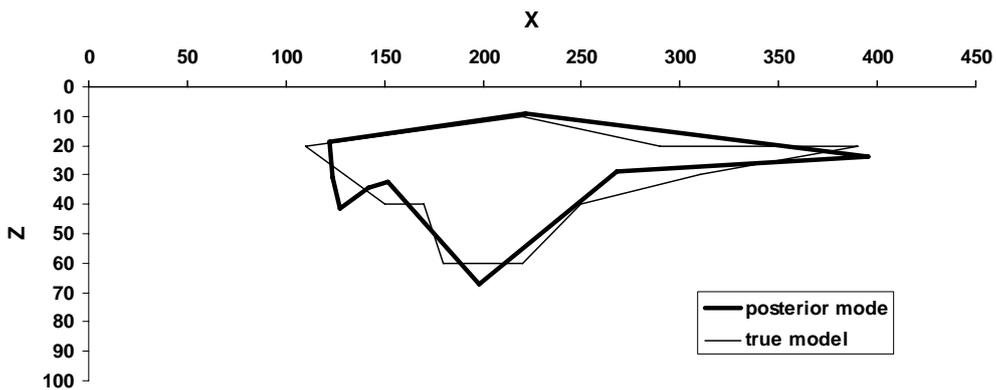

**Figure 12. Posterior mode for model $\mathcal{M}_9$, $\gamma = 1.2$.**

Results of for $\gamma = 2.0$ are shown in Figures 13 and 14. Comparing Figure 13 with Figure 6, there is now a downward shift of model probability towards polygons with a fewer number of vertices. The prior has over 70% probability for the triangle ($k = 3$), the simplest polygon. The highest posterior model probability now occurs at $k = 7$ (over 43%). This downward shift of model probability towards a fewer number of vertices is consistent with our expectation that a higher value in $\gamma$ imposes stronger restriction or penalty on the number of vertices. The posterior mode prediction of the most likely model $\mathcal{M}_7$ for $\gamma = 2.0$ is shown in Figure 14. Comparison between Figure 14 and Figure 8 seems to suggest that the overall agreement between the posterior prediction and the true shape is slightly better with $\gamma = 2.0$ than with $\gamma = 1.6$, even though $\mathcal{M}_7$ at $\gamma = 2.0$ has one fewer vertex than $\mathcal{M}_8$ at $\gamma = 1.6$. Thus $\gamma = 2.0$ is perhaps the preferred choice for this ill-conditioned problem, because it further enhances parsimony of the inferred model without losing accuracy of estimation.

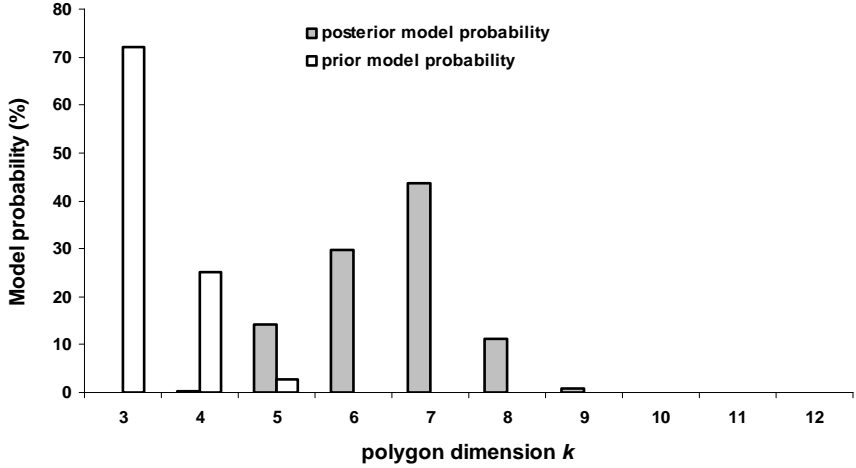

**Figure 13. Model probabilities predicted by MCMC posterior distribution and prior distribution.** $\gamma = 2.0$.

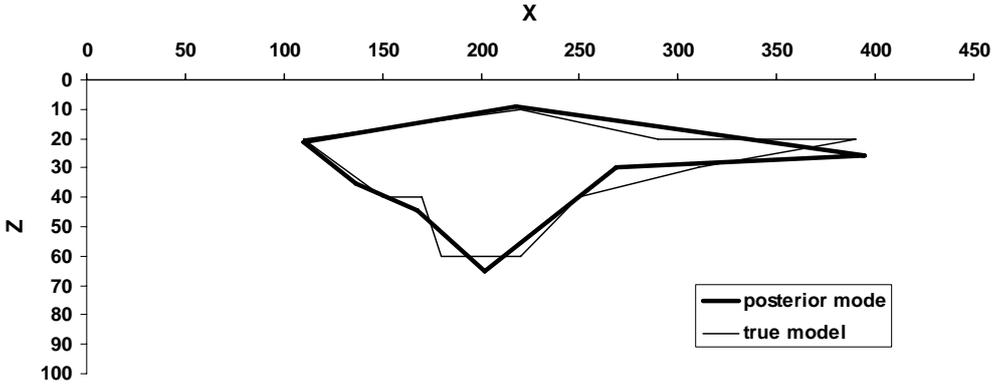

**Figure 14. Posterior mode for model $\mathcal{M}_7$, $\gamma = 2.0$.**

Numerical experiments show that if we use a uniform prior instead of the prior given by (8), i.e. if we assume the prior ratio is always unity (all admissible polygons are equally likely), then the rate of generating illegal polygons will be considerably higher, and convergence of the chain is much slower. Similar to the case of using prior (8) with $\gamma = 1.2$, the uniform prior leads to slightly poorer estimation on the polygon shape (in comparison with $\gamma = 1.6$), with a reduced efficiency and increased number of vertices for the most likely model. The challenge in real applications is of course to identify a good prior when the true model is not known.

## 5. Concluding remarks

The geophysical inverse problem of detecting the shape of an object underneath the earth surface, from the anomaly of gravity measured on the surface, is formulated as a trans-dimensional, statistical model inference problem, where the number of unknowns is one of the unknowns. A reversible jump Markov chain Monte Carlo (RJMCMC) algorithm for the polygonal earth model is proposed and implemented. Efficient within model and between model moves are proposed to increase the rate of admissible polygon samples. Numerical examples show RJMCMC is an efficient tool for solving such trans-dimensional inverse problems encountered in geophysics, which are typically ill-posed, non-linear and non-unique. Uncertainty in the solution is quantifiable and is actually an integral part of the statistical solution. The RJMCMC also appears to be parsimonious for the given prior, in the sense that among the models satisfactorily represent the true model, the model with fewer dimensions or number of parameters tend to have a higher posterior probability. Our sensitivity study shows a more restrictive prior (with a larger $\gamma$ value) can enhance parsimony of the inferred model without reducing accuracy of estimation for this ill-conditioned problem.

It is desirable and possible to devise more efficient Markov chain moves, both within and between models to reduce sampling rejection rate and to accelerate convergence to target distribution. It would be rather challenging to extend the present implementation to allow for multiple objects, where the number of objects is unknown, as well as the number of vertices, shape and location of each object. This extension is certainly not a trivial task and requires substantial research and programming work. Although in principle RJMCMC algorithm can be applied to such a problem, some theoretical and practical difficulties remain. For example, in the birth move to add a polygon, how can a new polygon be constructed and added to the existing polygons without overlapping and without greatly disturbing the solution to the forward problem (otherwise acceptance probability will be extremely low)? If we split an existing polygon for the birth move, there is infinitely more ways of splitting a polygon than removing a vertex of a polygon. What is the best way of splitting a polygon into two for the birth move? The death move is even more a challenge – removing a polygon is more likely to greatly disturb the solution of the forward problem than removing a vertex from a polygon. Perhaps two polygons getting too close to each other can be merged, and a death move occurs only in such occasions.

The proposed numerical algorithm can be readily adapted to similar but more realistic geophysical inverse applications, such as in three-dimensional object detection, provided that a proper parameterization is performed first and solutions for the forward problem can be computed efficiently. From statistical inversion point of view, the main difference between different geophysical inverse applications lies in the description of the forward problem – different likelihood functions resulting from different geophysics and measurement data are involved. As demonstrated by Bodin and Sambridge (2009), an adequate parameterization for the geophysical model should be regarded as an essential part of the inverse problem. The Bayesian MCMC is also ideal for combining different types of data, e.g. DC resistivity and gravity anomaly, to enable stronger inference with less uncertainty.

## Acknowledgement

The author would like to thank Gareth Peters, Pavel Shevchenko, Ross Sparks and Bob Anderssen for discussions and suggestions on this study, Malcolm Sambridge and an anonymous referee for reviewing the manuscript and providing helpful comments.